\documentclass[conference]{IEEEtran}
\usepackage{blindtext}
\usepackage{amsmath}
\usepackage{amssymb}
\usepackage{bm}
\usepackage{mdwmath}
\usepackage{xfrac}
\usepackage{graphicx}
\usepackage{xcolor}
\usepackage[caption=false,font=footnotesize,subrefformat=parens]{subfig}
\usepackage{makecell}
\usepackage{multirow}
\usepackage{acronym}
\usepackage[noadjust]{cite}
\usepackage{url}
\usepackage{algpseudocode}
\newcounter{MYalgorithmic}

\newcommand{\tabref}[1]{Table~\ref{#1}}

\newcommand{\upperroman}[1]{\uppercase\expandafter{\romannumeral#1}}

\begin{document}
\title{An Enhanced LMMSE Channel Estimation under High Speed Railway Scenarios}

\author{
	\IEEEauthorblockN{
		Qing~Tang,
		Hang~Long,
		Haojun~Yang,
		and Yuli Li}
	\IEEEauthorblockA{Wireless Signal Processing and Network (WSPN) Lab,\\
		Key Laboratory of Universal Wireless Communication, Ministry of Education,\\
		Beijing University of Posts and Telecommunications (BUPT), Beijing, 100876, China.}
		Email: qingtang@bupt.edu.cn}

\maketitle

\begin{abstract}
With the rapid deployment of the high speed railway (HSR), the wireless communication in HSR has been one of the indispensable scenarios in the fifth generation (5G) communications. In order to improve the performance of the orthogonal frequency division multiplexing (OFDM) system in the HSR scenarios, we propose an enhanced linear minimum mean square error channel estimation scheme based on multi-path Doppler frequency offset (DFO) estimation in this paper. The proposed scheme can estimate DFO of each path, and generate the frequency and time channel correlation more accurately, which can improve the accuracy of channel estimation in the HSR scenarios. Simulation results show that the proposed scheme can reduce the channel estimation error and achieve attractive gain in the HSR scenarios.
\end{abstract}
\IEEEpeerreviewmaketitle
\section{Introduction}
\label{sec:Introduction}
In the past few years, due to the large-scale deployment of high speed railway (HSR), the wireless communication in the HSR environment has been widely studied. In addition, the HSR communication has been incorporated into the fifth generation (5G) communications as a special scenario~\cite{my2,zhen3,final4}. 5G systems are expected to provide data rates up to 150 Mbps for a large number of users traveling at speeds up to 500 km/h in the HSR scenarios. However, the existing fourth generation (4G) long term evolution advanced (LTE-Advance) systems can only provide 2-4 Mbps data rate, so it is necessary to study the high-speed communication technology~\cite{my3,zhen2,final6}. \par
In order to meet the requirements of broad-band high data rate communication, the orthogonal frequency division multiplexing (OFDM) technology is also used for the HSR communication. However, in broad-band wireless communication systems, the channel is usually doubly selective~\cite{final1,zhen1,final2}. In order to equalize the received signal faded through the channel, the receiver must obtain the channel information of the whole channel. So the pilot based channel estimation method is widely used in OFDM system.\par
A large number of research on the channel estimation for the OFDM system have been propesed. The basis expansion model (BEM) is commonly used for the channel estimation in fast time-varying channels~\cite{my4,my5}, which is assumed that the channel rapidly changes in an OFDM symbol. It requires huge computational overhead due to approximate the channel impulse response by combinations of prescribed basis functions in the time domain. In order to reduce the computational complexity of the real systems, we assume the channel is quasi-static during an OFDM symbol. So the receiver can calculate the channel information at the pilot position and estimate the whole channel information in frequency domain. Several linear interpolation channel estimation algorithms are evaluated by~\cite{my6}, which can work well for a slowly fading channel. In addition, one of standard methods for channel estimation in frequency domain is linear minimum mean square error (LMMSE) channel estimation~\cite{my7}, which can effectively reduce the channel estimation error by using the frequency and time channel correlation. Nevertheless, in the channel with rapid time-variation, the above algorithm will bring serious performance loss due to inter carrier interference (ICI) and inaccurate channel correlation information.\par
Therefore, the objective of this paper is to propose a channel estimation scheme to improve the accuracy of channel estimation in the HSR scenarios. The main contributions of this paper includes:\par
\begin{itemize}
	\item We analyze ICI of the OFDM system, and derive the frequency and time channel correlation in the HSR scenarios.\par
	\item We propose a multi-path DFO estimation scheme, and analyze its complexity.\par
	\item We put forward an enhanced LMMSE (E-LMMSE) channel estimation scheme based on the proposed multi-path DFO estimation, improving the accuracy of channel estimation in HSR.\par
\end{itemize}
Simulation results show that our proposed algorithm can effectively reduce the channel estimation error and enhance the system performance in the HSR scenarios.\par
The rest of this paper is organized as follows. Section II introduces system model. In Section III, we put forward an E-LMMSE channel estimation scheme based on multi-path DFO estimation, and analyze its complexity. Section IV gives the simulation results and conclusions are drawn in Section V.\par

\textit{Notation:}  $ \mathbb{E}(\cdot) $ represents mathematical expectation, $ (\cdot)^\text{T} $ and $ (\cdot)^\text{*} $ denotes the transpose and the conjugate transpose of a matrix or vector, ${j}$ denotes plural unit.
\section{System Model}
\label{sec:System}
In this part, we introduce the two-tap HSR scenario and clarify the impact of ICI. Moreover, we derive channel correlation across sub-carriers and OFDM symbols, analyze the drawbacks of conventional channel estimation algorithms in the HSR scenarios.\par
\subsection{Scenario Description}
Most HSR are built in suburb between two cities. In order to have better network performance, a dedicated network is deployed along the HSR. To avoid frequent handovers, cell combination is applied: multiple remote radio heads (RRHs) are connected to one building baseband unit (BBU) with fiber. Then, from physical layer point of view, the cell coverage is extended significantly. \par
\begin{figure}[!t]
	\centering
	\includegraphics[width=0.45\textwidth]{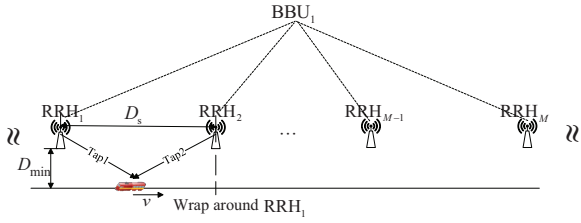}
	\caption{HSR two-tap channel model.}
	\label{Figure1}
\end{figure}

Fig.~\ref{Figure1} shows the two-tap HSR channel model defined in 3GPP~\cite{my1,final3}, where ${D_{\rm{s}}}$ is the distance between two neighbour RRH, ${{D_{\min }}}$ is the distance from RRH to railway. Assuming ${M}$ RRHs share the same cell identification and ${v}$ is the velocity of the train. Each RRH and user equipment (UE) on the train are equipped with two antennas, and UE is connected to two nearest RRHs. We denote the propagation paths from ${{\text{RR}}{{\text{H}}_{\text{1}}}}$ and ${{\text{RR}}{{\text{H}}_{\text{2}}}}$ to UE as Tap1 and Tap2. The two taps have independent DFOs, relative power and time delay. Moreover, each tap in this scenario is Rician fading. As ${D_{\text{min}}}$ is far less than ${D_{\text{s}}}$, angle of arrival (AOA) is very small so that DFO reach to more than 800Hz in most of the case.\par
\subsection{Signal Model}
The signal at the receiver can be denoted by
\begin{align}
\label{eqn_1}
& r\left( t \right) = s\left( t \right) + n\left( t \right) \notag \\
& = {\dfrac{1}{N}\sum\limits_{k = 0}^{N - 1} {{x_k}{e^{j2\pi k\Delta ft}}}}  \sum\limits_{q = 0}^1 {\tilde \sigma_q g( {t - {\tau _q}} ){e^{j2\pi {f_{d,q}}t}}} + n\left( t \right),
\end{align}
where $ {x_k} $ is the transmit signal, $ {N} $ is the inverse discrete Fourier transform (IDFT) size, $ {\Delta f} $ is the sub-carrier spacing, $ {\tilde \sigma {}_q} $ modeled as a complex random variable with uniformly distributed phase, $ {g\left( t \right)} $ is a rectangular window of the symbol length, $ {\tau _q} $ and $ {f_{d,q}} $ are time delay and DFO of the $ {q} $-th tap, ${n\left( t \right)} $ is Gaussian noise.
Convert $ {s\left( t \right)} $ to frequency domain, we can get the receive signal on $ {k} $-th sub-carrier of the $ {l} $-th OFDM symbol.
\begin{align}
\label{eqn_2}
S\left( {k,l} \right) &= \dfrac{1}{N}\sum\limits_{i = 0}^{N - 1} \bigg[ {x_i}\sum\limits_{q = 0}^1 \tilde \sigma {}_qG\left( {k - i - {F_{d,q}}} \right) \notag \\
&\mathrel{\phantom{=}} \times \; {e^{ - j\frac{{2\pi }}{N}{{\tilde \tau }_q}k}} \; {e^{j\frac{{2\pi }}{N}{F_{d,q}}l\left( {N + {N_{{\rm{CP}}}}} \right)}} \Big],
\end{align}
where ${{F_{d,q}} = {{{f_{d,q}}} \mathord{\left/
			{\vphantom {{{f_{d,q}}} {\Delta f}}} \right.
			\kern-\nulldelimiterspace} {\Delta f}}}$, ${\mathop {{\tau _q}}\limits^ \sim}$ is ${\tau _q}$ normalized by the sampling time ${T_s}$, ${N_{{\rm{CP}}}}$ is the length of cyclic prefix (CP), and
\begin{align}
\label{eqn_3}
G\left( k \right) = \sum\limits_{n = 0}^{N - 1} {{e^{ - j\frac{{2\pi k}}{N}n}}}  = {e^{ - j\pi k\left( {\frac{{N - 1}}{N}} \right)}}\frac{{\sin \left( {\pi k} \right)}}{{\sin \left( {\frac{{\pi k}}{N}} \right)}}.
\end{align}
Denote ${A_q}\left( {k,\Delta k} \right) = \frac{1}{N}\tilde \sigma {}_qG\left( {\Delta k - {F_{d,q}}} \right){e^{ - j\frac{{2\pi }}{N}{{\tilde \tau }_q}k}}$, then Eq.~\eqref{eqn_2} can be expressed as
\begin{align}
\label{eqn_4}
S\left( {k,l} \right) &= {x_k}\sum\limits_{q = 0}^1 {{A_q}\left( {k,0} \right){e^{j\frac{{2\pi }}{N}{F_{d,q}}l\left( {N + {N_{{\rm{CP}}}}} \right)}}} \notag \\
&\mathrel{\phantom{=}} + \sum\limits_{\begin{subarray}{l} 
	i=0, \\ 
	i \ne k 
	\end{subarray}}^{N - 1} {x_i} \Bigg[ {\sum\limits_{q = 0}^1 {{A_q}\left( {k,k - i} \right){e^{j\frac{{2\pi }}{N}{F_{d,q}}l\left( {N + {N_{{\rm{CP}}}}} \right)}}} } \Bigg] \notag \\
&= S_\text{fading}+S_\text{ICI}.
\end{align}
In Eq.~\eqref{eqn_4}, the expected signal on $ {k} $-th sub-carrier of the $ {l} $-th OFDM symbol is ${x_k}$. The second item ${S_\text{ICI}}$ corresponds to ICI from the other sub-carrier to the target sub-carrier, which does harm to the accuracy of the least square (LS) channel estimation. To simplify the analysis and without loss of generality, we assume the zero delay shifts for the two taps (one tap is due to the perfect time tracking and the other tap is assumed to be zero just for simplicity). Then the SIR ${\gamma}$ on the $ {k} $-th sub-carrier can be denoted as
\begin{align}
\label{eqn_5}
& \gamma \left( {{f_{d,0}},{f_{d,1}}} \right) =\lim_{L \rightarrow \infty} \dfrac{1}{L}\sum\limits_{l = 0}^{L - 1} \dfrac{\mathbb{E} \left[ S_\text{fading}\right]^2}{\mathbb{E} \left[ S_\text{ICI}\right]^2} \notag \\
& \mathop  \approx\limits^{(a)} \dfrac{{{{\left[ {\tilde \sigma {}_0G\left( { - {F_{d,0}}} \right)} \right]}^2} + {{\left[ {\tilde \sigma {}_1G\left( { - {F_{d,1}}} \right)} \right]}^2}}}{{\sum\limits_{\begin{subarray}{l} 
			i=0, \\ 
			i \ne k 
			\end{subarray}}^{N - 1} {{{\left[ {\tilde \sigma {}_0G\left( {k - i - {F_{d,0}}} \right)} \right]}^2} + {{\left[ {\tilde \sigma {}_1G\left( {k - i - {F_{d,1}}} \right)} \right]}^2}} }},
\end{align}
where ${a}$ denotes ${\mathop {{\tau _0}}\limits^ \sim   = \mathop {{\tau _1}}\limits^ \sim   = 0}$. According to Eq.~\eqref{eqn_5}, we can figure out ${\gamma}$ is around 20 dB in-between RRHs. Considering that the received power in-between RRHs is relatively small compared to the location close to the RRH, thus the performance would be primarily noise-limited instead of ICI-limited.\par
\label{sss_2}
\subsection{Channel Correlation}
We can divide ${S\left( {k,l} \right)}$ by the transmit signal ${x_k}$ to get the LS channel estimates
\begin{align}
\label{eqn_8}
 {H_\text{LS}}\left( {k,l} \right) & =\frac{{S\left( {k,l} \right)}}{{{x_k}}}=\sum\limits_{q = 0}^1 {{A_q}\left( {k,0} \right){e^{j\frac{{2\pi }}{N}{F_{d,q}}l\left( {N + {N_{{\rm{CP}}}}} \right)}}} \notag \\
&+ \sum\limits_{\begin{subarray}{l} 
	i=0, \\ 
	i \ne k 
	\end{subarray}}^{N - 1} {\frac{{{x_i}}}{{{x_k}}}} \Bigg[ {\sum\limits_{q = 0}^1 {{A_q}\left( {k,k - i} \right){e^{j\frac{{2\pi }}{N}{F_{d,q}}l\left( {N + {N_{{\rm{CP}}}}} \right)}}} } \Bigg] \notag \\
&= H\left( {k,l} \right) + w\left( {k,l} \right).
\end{align}
Due to the randomness of ${{{{x_i}} \mathord{\left/{\vphantom {{{x_i}} {{x_k}}}} \right.\kern-\nulldelimiterspace} {{x_k}}}}$, we can treat the second part (corresponding to ICI) of Eq.~\eqref{eqn_8} as noise.
For a given OFDM symbol, the correlation function across sub-carriers can be expressed as
\begin{align}
\label{eqn_10}
&\mathrel{\phantom{=}} R_{{\rm{H,F}}}^{{\rm{HSR}}}\left( {k - m} \right) = \mathbb{E}\left[ {H\left( {k,l} \right)H{{\left( {m,l} \right)}^*}} \right] \notag \\
&\approx \mathbb{E}\bigg[ {\sum\limits_{q = 0}^1 {{A_q}\left( {k,0} \right){A_q}\left( {m,0} \right)} } \bigg] \notag \\
&= \dfrac{1}{{{N^2}}} \sum\limits_{q = 0}^1 {{{\left| {\tilde \sigma {}_q} \right|}^2}{{\left| {G\left( { - {F_{d,q}}} \right)} \right|}^2}{e^{ - j\frac{{2\pi }}{N}{{\tilde \tau }_q}\left( {k - m} \right)}}} .
\end{align}
For a given sub-carrier, the correlation function across OFDM symbols can be expressed as
\begin{align}
\label{eqn_11}
&\mathrel{\phantom{=}} R_{{\rm{H,T}}}^{{\rm{HSR}}}\left( {x - l} \right) = \mathbb{E} \left[ {H\left( {k,x} \right)H{{\left( {k,l} \right)}^*}} \right] \notag \\
&\approx \mathbb{E} \bigg[ {\sum\limits_{q = 0}^1 {{{\left| {{A_q}\left( {k,0} \right)} \right|}^2}{e^{j\frac{{2\pi }}{N}{F_{d,q}}\left( {x - l} \right)\left( {N + {N_{{\rm{CP}}}}} \right)}}} } \bigg] \notag \\
&= \sum\limits_{q = 0}^1 {\frac{1}{{{N^2}}}{{\left| {\tilde \sigma {}_q} \right|}^2}{{\left| {G\left( { - {F_{d,q}}} \right)} \right|}^2}{e^{j2\pi {f_{d,q}}\left( {x - l} \right)\left( {N + {N_{{\rm{CP}}}}} \right){T_s}}}}  .
\end{align}

\subsection{Drawbacks of Conventional Channel Estimation Algorithms in HSR}
The linear interpolation algorithm assumes that the channel changes slowly in time domain and frequency domain, so that it uses the linear interpolation function to get the frequency response of the whole channel from the pilot position. As shown in~\eqref{eqn_10} and~\eqref{eqn_11}, the channel does not change linearly, which brings performance loss in HSR scenarios.\par
The conventional LMMSE channel estimation is applied to 2-D isotropic scattering, the frequency domain and time domain autocorrelation function can be expressed as
\begin{align}
\label{eqn_6}
{R_{{\rm{H,F}}}^{{\rm{legacy}}}}\left( {k\Delta f} \right) = \sum\limits_{q = 0}^{L - 1} {{{\left| {\tilde \sigma {}_q} \right|}^2}{e^{ - j2\pi k\Delta f{{\tilde \tau }_q}}}},
\end{align}
\begin{align}
\label{eqn_7}
{R_{{\rm{H,T}}}^{{\rm{legacy}}}}\left( {lT} \right) = {J_0}\left( {2\pi l{f_D}T} \right),
\end{align}
where ${L}$ is the total taps in the channel model, ${T}$ is the duration of an OFDM symbol, ${f_D}$ is the maximum DFO, ${{J_0}\left(  \cdot  \right)}$ is the 0-th Bessel function of the first kind. The channel correlation ${R_{{\rm{H,F}}}^{{\rm{HSR}}}\left( {k - m} \right)}$ and ${R_{{\rm{H,T}}}^{{\rm{HSR}}}\left( {x - l} \right)}$ in HSR scenarios are much different from ${R_{{\rm{H,F}}}^{{\rm{legacy}}}}$ and ${R_{{\rm{H,T}}}^{{\rm{legacy}}}}$. When conventional LMMSE channel estimation are directly using in HSR scenarios, the system must have some performance loss. This is mainly because the time domain and frequency domain wiener filtering use inaccurate correlation function. Therefore, motivated by the above facts, we propose an E-LMMSE channel estimation to improve the system performance for HSR scenarios in Section III.
\section{E-LMMSE Channel Estimation Algorithm Based On Multipath DFO Estimation}
In this part, we propose a novel multi-path DFO estimation algorithm, then generate accurate channel correlation across sub-carriers and OFDM symbols, which can be used in E-LMMSE to improve the accuracy of channel estimation in HSR scenarios. Moreover, we compare the complexity of the proposed scheme with the exhaustive search (ES) estimation algorithm in~\cite{my8}.\par
\subsection{Multipath DFO Estimation Algorithm}
From Eq.~\eqref{eqn_10} and Eq.~\eqref{eqn_11}, the receiver should get multi-path DFO to calculation accurate channel correlation in HSR scenarios. In~\cite{my8}, an ES estimation algorithm for multi-path DFO estimation was proposed, but its complexity is too high due to exhaustive search. Here, we propose a novel method to estimate multi-path DFO.\par
All the channel observations on pilot index of a OFDM symbol can be estimated by Eq.~\eqref{eqn_8}, we group them by their OFDM symbol index. Denote ${{l_1},{l_2}, \ldots {l_n}}$ as the indices of the OFDM symbols containing pilots, where ${n}$ is the total number of such OFDM symbol. So we can denote the column vector ${
{\bf{H}}_p^i,i = 1,2, \ldots n}$ as all the channel observations on pilot index of the ${{l_i}}$-th OFDM symbol. And the channel noise on the pilot index of the ${{l_i}}$-th OFDM symbol can be denoted as the column vector ${{\bf{W}}_p^i}$.  Moreover, define a matrix ${{{\bf{D}}_p}}$ related with delay and carrier index.
\begin{align}
\label{eqn_12}
{{\bf{D}}_p} = \left[ {\begin{array}{*{20}{c}}
	{\begin{array}{*{20}{c}}
		{{e^{ - j\frac{{2\pi }}{N}{{\tilde \tau }_0}{k_1}}}}\\
		{\begin{array}{*{20}{c}}
			\vdots \\
			{{e^{ - j\frac{{2\pi }}{N}{{\tilde \tau }_0}{k_p}}}}
			\end{array}}\\
		\vdots \\
		{{e^{ - j\frac{{2\pi }}{N}{{\tilde \tau }_0}{k_m}}}}
		\end{array}}&{\begin{array}{*{20}{c}}
		{{e^{ - j\frac{{2\pi }}{N}{{\tilde \tau }_1}{k_1}}}}\\
		{\begin{array}{*{20}{c}}
			\vdots \\
			{{e^{ - j\frac{{2\pi }}{N}{{\tilde \tau }_1}{k_p}}}}
			\end{array}}\\
		\vdots \\
		{{e^{ - j\frac{{2\pi }}{N}{{\tilde \tau }_1}{k_m}}}}
		\end{array}}
	\end{array}} \right],
\end{align}
where ${m}$ represents the total number of carriers containing pilots on an OFDM symbol, ${k_p}$ represents the carrier index with pilot. Then the relation of them in the frequency domain can be classified as
\begin{align}
\label{eqn_13}
{\bf{H}}_p^i = {{\bf{D}}_p}{\bf{X}}_p^i + {\bf{W}}_p^i,
\end{align}
where 
${{\bf{X}}_p^i = \left[ {X_1^i,X_2^i} \right]^\text{T}}$,

\begin{align}
\label{eqn_14}
{X_q^i = \frac{1}{N}\tilde \sigma {}_qG\left( { - {F_{d,q}}} \right){e^{j\frac{{2\pi }}{N}{F_{d,q}}l\left( {N + {N_{{\rm{CP}}}}} \right)}}\quad q = 0,1},
\end{align}
then using the LS method, we can obtain the estimated value of ${{\bf{X}}_p^i}$ 
\begin{align}
\label{eqn_15}
{\bf{Z}}_p^i = {\left( {{{\bf{D}}_p}} \right)^{ - 1}}{\bf{H}}_p^i + {\left( {{{\bf{D}}_p}} \right)^{ - 1}}{\bf{W}}_p^i,
\end{align}
where ${{\bf{Z}}_p^i = \left[ {Z_1^i,Z_2^i} \right]^\text{T}}$. Considering the pilots in from the ${l_1}$-th to ${l_n}$-th OFDM symbols, we will render
\begin{align}
\label{eqn_16}
{{\bf{Z}}_p} = {\left( {{{\bf{D}}_p}} \right)^{ - 1}}{{\bf{H}}_p} + {\left( {{{\bf{D}}_p}} \right)^{ - 1}}{{\bf{W}}_p},
\end{align}
where ${{{\bf{Z}}_p} = \left[ {{\bf{Z}}_p^1, \cdots ,{\bf{Z}}_p^n} \right]}$, ${{{\bf{H}}_p} = \left[ {{\bf{H}}_p^1, \cdots ,{\bf{H}}_p^n} \right]}$, ${{{\bf{W}}_p} = \left[ {{\bf{W}}_p^1, \cdots ,{\bf{W}}_p^n} \right]}$.
${{{\bf{W}}_p}}$ is assumed to be additive background white Gaussian noise in order to simplify the problem. Moreover, the noise is assumed to be independent to matrix ${{{\bf{D}}_p}}$. So we can think ${
{\left( {{{\bf{D}}_p}} \right)^{ - 1}}{{\bf{W}}_p}}$ is identical to ${{{\bf{W}}_p}}$ in statistical sense. Therefore, ${{\bf{Z}}_p^i}$ is the estimated value of ${{\bf{X}}_p^i}$ in ${l_i}$-th OFDM symbol. For any two columns ${{\bf{Z}}_p^i}$ and ${{\bf{Z}}_p^j}$ in ${{{\bf{Z}}_p}}$, we can get the multi-path DFO estimation by Eq.~\eqref{eqn_17}.
\begin{align}
\label{eqn_17}
\mathop {{f_{d,q}}}\limits^ \sim   = \frac{{{\rm{angle}}\left( {{{Z_q^j} \mathord{\left/
					{\vphantom {{Z_q^j} {Z_q^i}}} \right.
					\kern-\nulldelimiterspace} {Z_q^i}}} \right)N\Delta f}}{{2\pi \left( {N + {N_{{\rm{CP}}}}} \right)\left( {{l_j} - {l_i}} \right)}}\quad q = 1,2,
\end{align}
where ${\mathop {{f_{d,0}}}\limits^ \sim}$ is the estimated value of ${f_{d,0}}$, ${\mathop{{f_{d,1}}}\limits^ \sim}$ is the estimated value of ${f_{d,1}}$.
Do Eq.~\eqref{eqn_17} to a number of columns in ${{{\bf{Z}}_p}}$, then take average of the estimated values ${\mathop {{f_{d,0}}}\limits^ \sim}$ and ${\mathop{{f_{d,1}}}\limits^ \sim}$, we can get a more accurate estimated value.
\subsection{E-LMMSE Channel Estimation Algorithm Based On Multipath DFO Estimation}
LMMSE channel estimation requires the use of channel correlation, so Eq.~\eqref{eqn_6} and Eq.~\eqref{eqn_7} are used in the conventional LMMSE channel estimation, while it is not suitable for HSR scenarios. Therefore, we propose an E-LMMSE channel estimation algorithm based on multi-path DFO estimation.
The receiver can calculate the correlation function Eq.~\eqref{eqn_10} and Eq.~\eqref{eqn_11} by using ${\mathop {{f_{d,0}}}\limits^ \sim}$ and ${\mathop{{f_{d,1}}}\limits^ \sim}$, and generate channel correlation matrix ${{{\bf{R}}_{{\rm{H,F}}}^{{\text{HSR}}}}}$ and ${{{\bf{R}}_{{\rm{H,T}}}^{{\text{HSR}}}}}$ needed by Wiener filter~\cite{my9,final5}. Then the receiver can obtain frequency domain wiener filter matrix ${{\bf{W}}_{{\text{F}}}^{{\text{HSR}}}}$ and time domain wiener filter matrix ${{\bf{W}}_{{\text{T}}}^{{\text{HSR}}}}$ by
\begin{align}
\label{eqn_18}
{{\bf{W}}_{{\text{F}}}^{{\text{HSR}}}} = {{\bf{R}}_{{\text{H,F}}}^{{\text{HSR}}}}{\left[ {{{\bf{R}}_{{\text{H,F}}}^{{\text{HSR}}}} + \frac{{{\sigma ^2}}}{{\mathbb{E}\left[ {{{\left| {{x_p}} \right|}^2}} \right]}}{\bf{I}}} \right]^{ - 1}},
\end{align}
\begin{align}
\label{eqn_20}
{{\bf{W}}_{{\text{T}}}^{{\text{HSR}}}} = {{\bf{R}}_{{\text{H,T}}}^{{\text{HSR}}}}{\left[ {{{\bf{R}}_{{\text{H,T}}}^{{\text{HSR}}}} + \frac{{{\sigma ^2}}}{{\mathbb{E}\left[ {{{\left| {{x_p}} \right|}^2}} \right]}}{\bf{I}}} \right]^{ - 1}},
\end{align}
where ${{x_p}}$ is pilot information. Finally, Wiener filter is used to smoothing the LS channel estimate,
\begin{align}
\label{eqn_19}
{{\bf{H}}_{{\rm{Est}}}} = {{\bf{W}}_{{\text{T}}}^{{\text{HSR}}}}{{\bf{W}}_{{\text{F}}}^{{\text{HSR}}}}{{\bf{H}}_{{\text{LS}}}},
\end{align}
where ${{\bf{H}}_{{\rm{Est}}}}$ is frequency response of channel.
\subsection{Complexity Analysis}
We compare the complexity between the proposed DFO estimation algorithm with the ES estimation algorithm. The main calculation process of the proposed DFO estimation is calculating inverse of the matrix ${{{\bf{D}}_p}}$, which takes ${8m}$ multiplications. Then the operation ${{\left( {{{\bf{D}}_p}} \right)^{ - 1}}{{\bf{W}}_p}}$ takes ${2m}$ multiplications. So calculating ${n}$ OFDM symbols in total, it calls for ${10mn}$ multiplications. On the other hand, assuming maximum search range and search step size in the ES estimation algorithm are ${f_{\max }}$ and ${\delta}$, then it calls for ${{{m{f_{\max }}\left( {8n + 6} \right)} \mathord{\left/
			{\vphantom {{m{f_{\max }}\left( {8n + 6} \right)} \delta }} \right.
			\kern-\nulldelimiterspace} \delta }}$ multiplications. For example, in LTE-A downlink system, the proposed DFO estimation algorithm takes 80 multiplications in one resource block (RB), while the ES estimation algorithm takes 34200 multiplications. Therefore it has reduced 400 times multiplications than the ES estimation algorithm. The complexity of two algorithms are illustrated in \tabref{table_1}.
\begin{table}[!t]
	\renewcommand{\arraystretch}{1.2}
	\setlength{\extrarowheight}{1pt}
	\centering
	\caption{The Number of Multiplications for Two DFO Estimation Methods.}
	\begin{tabular}{| l | l |}
		\hline
		\textbf{Algorithm} & \textbf{Multiplications} \\
		\hline
		\text{Proposed DFO estimation} & ${10mn}$ \\
		\hline
		\text{ES estimation} & ${{{m{f_{\max }}\left( {8n + 6} \right)} \mathord{\left/
					{\vphantom {{m{f_{\max }}\left( {8n + 6} \right)} \delta }} \right.
					\kern-\nulldelimiterspace} \delta }}$ \\
		\hline
	\end{tabular}
	\label{table_1}
\end{table}
\section{Simulation Results}
In this section, we contrast the proposed DFO estimation algorithm with the ES estimation algorithm in terms of estimation error. In addition, the performance between E-LMMSE channel estimation and the conventional channel estimation methods will be compared. The simulation parameters are shown in \tabref{table_2}.\par
\begin{table}[!t]
	\renewcommand{\arraystretch}{1.2}
	\setlength{\extrarowheight}{1pt}
	\centering
	\caption{Simulation Parameters.}
	\begin{tabular}{| l | l |}
		\hline
		\textbf{Parameter} & \textbf{Assumption} \\
		\hline
		Antenna configuration & ${2 \times 2}$ \\
		\hline
		Carrier frequency & 2.6 GHz \\
		\hline
		Train speed ${v}$ & 350 km/h \\
		\hline
		HSR scenario & the two-tap scenario \\
		\hline
		FFT size ${N}$ & 1024 \\
		\hline
		RB  & 50 \\
		\hline
		System bandwidth  & 10 MHz \\
		\hline
		Modulation  & 16QAM \\
		\hline
	\end{tabular}
	\label{table_2}
\end{table}
\begin{figure}[!t]
	\centering
	\includegraphics[width=0.45\textwidth]{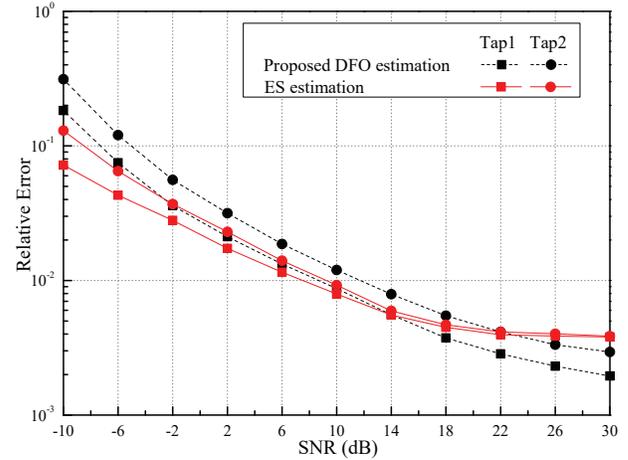}
	\caption{Relative error of two DFO estimation algorithms at the 3 dB power-difference position.}
	\label{Figure3}
\end{figure}
Fig.~\ref{Figure3} shows the relative error normalized by the maximum DFO (842 Hz) of two DFO estimation methods. The simulation was conducted in the fixed position closing to ${{\text{RR}}{{\text{H}}_{\text{1}}}}$, where the received power of Tap1 is twice than the received power of Tap2 (3 dB power-difference position). As a result, the DFO estimation of Tap1 is more accurate. With the increasing of SNR, the performance of the proposed DFO estimation is better than the ES algorithm, that is because search step size ${\delta}$ in the ES algorithm is fixed to 2. Besides, two DFO estimation algorithms have an error flat, which is due to the existence of ICI in the HSR scenarios. As mentioned in Section~\ref{sss_2}, the ICI is about -20 dB.\par
\begin{figure}[!t]
	\centering
	\includegraphics[width=0.45\textwidth]{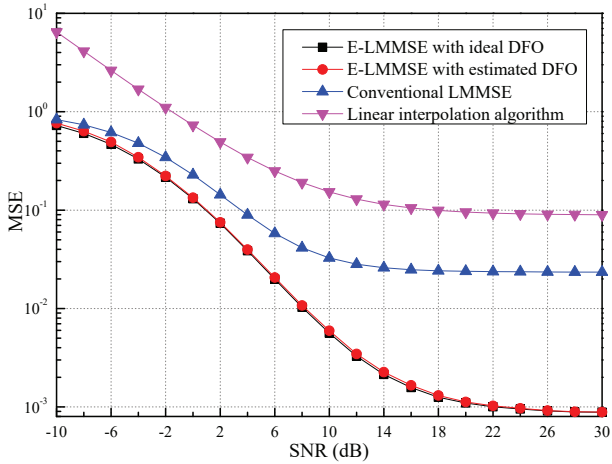}
	\caption{MSE of several channel estimation methods at the 3 dB power-difference position.}
	\label{Figure4}
\end{figure}
\begin{figure}[!t]
	\centering
	\includegraphics[width=0.45\textwidth]{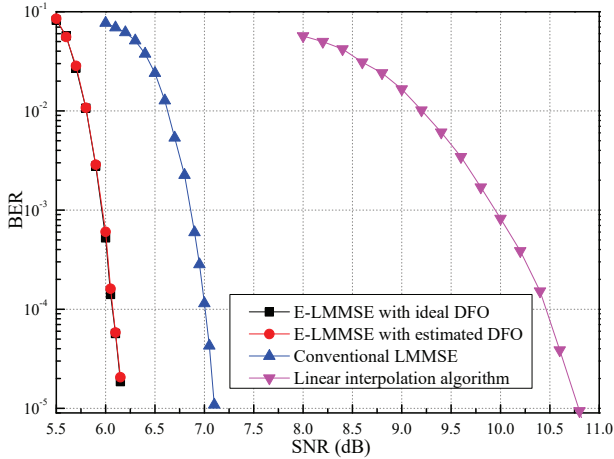}
	\caption{BER of several channel estimation methods at the 3 dB power-difference position.}
	\label{Figure5}
\end{figure}
\begin{figure}[!t]
	\centering
	\includegraphics[width=0.45\textwidth]{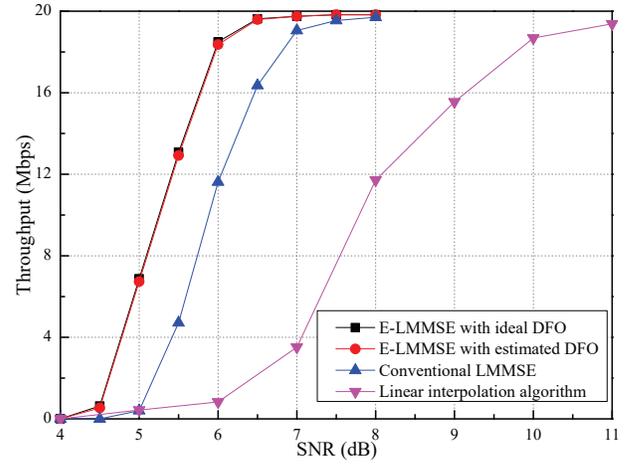}
	\caption{Average TP from one RRH to the next one.}
	\label{Figure6}
\end{figure}

Fig.~\ref{Figure4} compares the channel estimation MSE of several channel estimation algorithms. Observe that the performance of E-LMMSE with estimated DFO is similar to E-LMMSE with ideal DFO channel estimation, which means the estimation error of the proposed DFO estimation scheme is small enough and does not effect the performance of E-LMMSE. Besides, the channel estimation of E-LMMSE is much precise than the conventional LMMSE channel estimation and linear interpolation algorithm. For example, when SNR
is 30 dB, our E-LMMSE method achieves more than 10 dB MSE gain than the conventional methods. But it still has an error limit due to the ICI. Fig.~\ref{Figure5} compares the bit error rate (BER) of several channel estimation methods at the 3 dB power-difference position. E-LMMSE channel estimation also gets a great improvement than the conventional methods under the performance of BER. Fig.~\ref{Figure6} compares the average throughput (TP) of several channel estimation methods from one RRH to the next one, and the performance of E-LMMSE is attractive than the conventional channel estimation algorithms.
\section{Conclusion}
\label{sec:Conclusion}
In this paper, in order to improve the performance of OFDM systems in HSR scenarios, we proposed an E-LMMSE channel estimation scheme based on multi-path DFO estimation. The proposed method generates the frequency and time channel correlation more precisely by estimating multi-path DFO, which can improve the accuracy of channel estimation in HSR scenarios. Simulation results show that the proposed method can accurately estimate multi-path DFO, and reduce the channel estimation error. Moreover, it also achieves attractive gain under the performance of BER and TP in HSR systems.

\section*{Acknowledgment}
This work is supported by the National Key Scientific Instrument and Equipment Development Project 2013YQ20060706, the National High Technology Research and Development Program of China under Grant 2014AA01A705, the National Key Technology R\&D Program of China under Grant 2015ZX03002009-004 and China Unicom.

%


\end{document}